\documentclass[appendixfloats]{emulateapj}
\usepackage{longtable, hyperref, graphicx, subfigure}
\usepackage[fleqn]{amsmath}
\usepackage{lmodern}
\usepackage[T1]{fontenc}
\usepackage{url}
\newcounter{column_number}
\newcommand{\Msun}{\ifmmode {M_{\odot}}\else${M_{\odot}}$\fi}
\newcommand{\Rsun}{\ifmmode {R_{\odot}}\else${R_{\odot}}$\fi}
\newcommand{\lapprox }{{\lower0.8ex\hbox{$\buildrel <\over\sim$}}}
\newcommand{\gapprox }{{\lower0.8ex\hbox{$\buildrel >\over\sim$}}}

\def\asec{\ifmmode^{\prime\prime}\else$^{\prime\prime}$\fi}

\def\ergs{$\mathrm{erg}$~$\mathrm{s}^{-1}$}

\def\ROSAT{\textit{ROSAT}}
\def\XMM{\textit{XMM-Newton}}
\def\CHANDRA{\textit{Chandra}}
\def\LX{$L_{\mathrm{X}}$}
\def\LL{$L_{\mathrm{X}}/L_{\mathrm{bol}}$}
\def\Lbol{$L_{\mathrm{bol}}$}

\def\Pmem{$P_{\mathrm{mem}}$}

\slugcomment{DRAFT \today}
\shorttitle{M37 and the Evolution of X-ray Emission in Low-Mass Stars}
\shortauthors{N\'u\~nez \& Ag{\"u}eros}

\begin{document}

\title{The X-ray luminosity function of M37 and the evolution of coronal activity in low-mass stars} 
\author{Alejandro N\'u\~nez\altaffilmark{1} and
Marcel~A.~Ag{\"u}eros\altaffilmark{1}
}

\altaffiltext{1}{Department of Astronomy, Columbia University, 550 West 120th Street, New York, NY 10027, USA} 

\begin{abstract}
We use a 440.5~ks \CHANDRA\ observation of the $\approx$500-Myr-old open cluster M37 to derive the X-ray luminosity functions of its $\leq$1.2~\Msun\ stars. Combining detections of 162 M37 members with upper limits for 160 non-detections, we find that its G, K, and M stars have a similar median (0.5$-$7 keV) X-ray luminosity \LX~$=10^{29.0}$ \ergs, whereas the \LX-to-bolometric-luminosity ratio (\LL) indicates that M stars are more active than G and K stars by $\approx$1 order of magnitude at 500 Myr.  To characterize the evolution of magnetic activity in low-mass stars over their first $\approx$600 Myr, we consolidate X-ray and optical data from the literature for stars in six other open clusters: from youngest to oldest, the Orion Nebula Cluster (ONC), NGC~2547, NGC~2516, the Pleiades, NGC~6475, and the Hyades. For these, we homogenize the conversion of instrumental count rates to \LX\ by applying the same one-temperature emission model as for M37, and obtain masses using the same empirical mass-absolute magnitude relation (except for the ONC). We find that for G and K stars X-ray activity decreases $\approx$2 orders of magnitude over their first 600 Myr, and for M stars, $\approx$1.5. The decay rate of the median \LX\ follows the relation \LX~$\propto~t^b$, where $b=-0.61\pm0.12$ for G, $-0.82\pm0.16$ for K, and $-0.40\pm0.17$ for M stars. In \LL\ space, the slopes are $-0.68\pm0.12$, $-0.81\pm0.19$, and $-0.61\pm0.12$, respectively. These results suggest that for low-mass stars the age-activity relation steepens after $\approx$625 Myr, consistent with the faster decay in activity observed in solar analogs at $t>1$ Gyr.
\end{abstract}


\section{Introduction}
In low-mass stars ($\lapprox$1~\Msun), X rays originate in a magnetically heated corona, and serve as a proxy for the strength of the magnetic dynamo\footnote{For a recent review of the connection between stellar activity and coronal heating, see \citet{Testa2015}.}. The X-ray luminosity, \LX, is correlated with age and rotation \citep[e.g.,][]{palla81, pizzo03}, and decreases as low-mass stars spin down because of the loss of angular momentum through magnetized winds. Calibrating the evolution of \LX\ is key to quantifying the interplay between stellar rotation and magnetic fields, and ultimately to uncovering the still-mysterious processes responsible for these fields.

Observations indicate that \LX\ does not decay smoothly with age ($t$). Surveys of solar-type stars in the Pleiades ($t \approx 125$ Myr) and the Orion Nebula Cluster (ONC; 1$-$10 Myr) concluded that \LX\ falls off relatively slowly early in a star's life: \LX\ $\propto t^{-0.76}$ \citep{queloz98,preibisch2005a}. Because there are few accessible $t\ \gapprox\ 200$ Myr clusters, constraints are harder to come by for older stars, but from observations of five solar analogs,  \citet{guedel1997} determined that \LX~$\propto t^{-1.5}$ for $t > 1$ Gyr, as did \citet{giardino2008} from their survey of the $\approx$1.5-Gyr-old cluster NGC 752. Core-envelope decoupling or a change in the magnetic field topology are the commonly invoked explanations for this sharp drop off in \LX\ \citep[e.g.,][]{kawaler1988,Krishnamurthi1997}, but it remains poorly understood. Interestingly, chromospheric activity, another proxy for magnetic field strength, appears to evolve differently \citep[e.g.,][]{Jackson2010, Douglas14} but may suffer a similarly steep decline for $t>1$ Gyr \citep{pace2004}.  

To determine the evolution of \LX, we need more $\gapprox$200-Myr-old open clusters with well-characterized cumulative X-ray luminosity functions (XLFs). In \citet[][hereafter Paper I]{Nunez2015}, we described our 440.5 ks {\it Chandra} observation of M37 (NGC 2099), a rich, $\approx$500-Myr-old cluster at a distance of 1490$\pm$120 pc \citep{hartman08b}. We combined the photometry compiled by \citet[hereafter HA08]{Hartman2008a} and distance from the cluster core to generate membership probabilities (\Pmem) for cluster stars. Our final catalog included 561 X-ray sources with optical counterparts, 278 of which had \Pmem~$\geq 0.2$. Here, we add to these detections \LX\ upper limit (UL) measurements for undetected members to determine the XLFs for M37's G, K, and M stars. We also compute bolometric luminosities (\Lbol), and use these to determine the \LL\ functions (LLFs) for these stars, thereby establishing M37 as the benchmark 500-Myr-old cluster for studies of the evolution of X-ray emission in low-mass stars.

In Section 2, we describe our optical and X-ray data for M37, outline how we assign membership thresholds for inclusion in our analysis, and calculate \LX\ ULs for undetected sources. In Section 3, we construct the XLFs and LLFs and discuss the impact of our upper limits and of binaries on these functions. In Section 4, we first homogenize the \LX, \Lbol, and masses of stars in six other clusters ranging in age from 6 to 625 Myr. We then examine the evolution of the XLFs and LLFs for GKM stars over $\approx$600 Myr. We present our conclusions in Section 5.  

\section{Characterizing Low-Mass Stars in M37}\label{lowmass}
\subsection{Setting the Thresholds for Cluster Membership}
HA08 obtained {\it gri} images of a 24$\arcmin$$\times$24$\arcmin$ area centered on M37 with Megacam on the 6.5-m MMT telescope at the MMT Observatory, AZ. In Paper I, we used this photometry and the distance from the cluster core for $\approx$16,800 HA08 objects to identify cluster members. Each star was assigned a probability of being a single member ($P_{\mathrm{s}}$), a likely binary member ($P_{\mathrm{b}}$), or a field star ($P_{\mathrm{f}}$), with $P_{\mathrm{s}}$ + $P_{\mathrm{b}} \equiv$ \Pmem\ and $P_{\mathrm{s}}$ + $P_{\mathrm{b}}$ + $P_{\mathrm{f}}$ = 1 for each star. We identified 1643 stars with \Pmem\ $\geq$ 0.2, which we used as the \Pmem\ cutoff for cluster membership (see section 3.3 and appendix A of Paper I). 

Using such a low \Pmem\ threshold increases the likelihood of field-star contamination: for example, only 20$-$30\% of the stars in the $0.2\leq$\Pmem$<0.3$ bin should be {\it bona fide} cluster members. However, because M37 stars are much more likely to be bright X-ray emitters than their older field-star cousins, we consider all X-ray emitters with \Pmem\ $>$ 0.2 to be cluster members and now assign these stars \Pmem\ = 1.0. 

In effect, this redistributes absolute \Pmem\ points from the non-detections to the X-ray emitters by the total quantity $q=\sum_{i} (1-$ \Pmem$_{,i})$, where $i$ is the number of X-ray emitters in each \Pmem\ bin ($0.2\leq $ \Pmem $<0.3$, $0.3\leq $ \Pmem $<0.4$, and so on). We therefore also decrease the \Pmem\ values of non-detections by subtracting the quantity $q/N$ from their original \Pmem, where $N$ is the number of non-detections in each \Pmem\ bin.

While we could simply use the original \Pmem~$\geq 0.2$ threshold for non-detections, there is no way to distinguish {\it bona fide} members that are X-ray-undetected from field contaminants in low \Pmem\ bins. We therefore apply a more conservative \Pmem~$\geq 0.7$ cutoff to minimize the risk of biased results. Finally, we note that for stars outside the field-of-view of our \CHANDRA\ observation \Pmem~$\leq 0.4$, so that the completeness of our sample is not affected by excluding these stars.

\subsection{Assigning Masses and Estimating \Lbol}\label{masses}
To estimate masses for M37 stars, we first used the $r$ photometry of HA08, along with the cluster reddening $E$($B-V$) = 0.227 and distance of 1490 pc obtained by these authors, to calculate an absolute $r$ magnitude $M_r$ for each member. We then applied the $M_r$-mass relation of \citet{Kraus2007}, who generated empirical spectral energy distributions for B8-L0 stars that are calibrated using the 600-Myr-old Praesepe cluster and Sloan Digital Sky Survey \citep[SDSS,][]{york00} $ugriz$ and Two Micron All Sky Survey \citep[2MASS,][]{2mass} $JHK$ photometry (see section 4.1.1 of Paper I). 

We estimate \Lbol\ for M37 members by first using the $M_r$-effective temperature ($T_{eff}$) relation of \citet{Kraus2007}. After obtaining $T_{eff}$ for each star, we use the corresponding bolometric correction in the \citet{Girardi04} tables, which we tailor to the SDSS filter system. This allows us to calculate bolometric magnitudes and luminosities, the latter by again using the distance to the cluster of 1490 pc.

There are 118 M37 stars with \Pmem~$\geq 0.7$ and masses 0.8$-$1.2 \Msun, 125 with masses 0.6$-$0.8 \Msun, and 79 with masses 0.1$-$0.6 \Msun, corresponding approximately to G, K, and M stars, respectively \citep{Cram1989}. Table~\ref{tbl:M37} lists all low-mass M37 stars with \Pmem~$\geq 0.2$, although only those with \Pmem~$\geq 0.7$ were used in our study. Column 1 is the source ID of the optical object from HA08; Column 2 is the recalculated \Pmem; Column 3 is a binary flag set to 1 if the object is likely a binary (see appendix A in Paper I) and 0 otherwise; Column 4 is mass; Column 5 is  \Lbol; Column 6 is a detection flag set to 1 if the object has an X-ray detection and 0 otherwise (i.e., a non-detection); Column 7 is the \LX\ of detections or the \LX\ ULs of non-detections (see discussion below).

\tabcolsep=0.08cm
\begin{deluxetable}{@{}c@{}c@{}c@{}c@{}c@{}c@{}c@{}}
\tabletypesize{\scriptsize}
\tablecaption{M37 Low-Mass Members in the Field of View of \CHANDRA \label{tbl:M37}}

\tablehead{
\colhead{ID\tablenotemark{a}} & \colhead{\Pmem\tablenotemark{b}} & \colhead{Bin.} & \colhead{Mass} & \colhead{log(\Lbol)\tablenotemark{d}} & \colhead{Det.} & \colhead{log(\LX)\tablenotemark{f}}\\  
\colhead{} & \colhead{} & \colhead{flag\tablenotemark{c}} & \colhead{(\Msun)} & \colhead{(\ergs)} & \colhead{flag\tablenotemark{e}} & \colhead{(\ergs)}\\[-0.1 in]
}
\startdata
30118 & 0.48 & 0 & 0.61 & 32.55 & 0 & 29.63\\
40031 & 0.42 & 0 & 0.98 & 33.26 & 0 & 29.52\\
40039 & 0.41 & 0 & 0.88 & 33.12 & 0 & 29.62\\
40097 & 0.41 & 0 & 0.66 & 32.72 & 0 & 29.67\\
40103 & 0.60 & 0 & 0.63 & 32.66 & 0 & 29.57
\enddata
\tablenotetext{a}{Source ID from HA08.}
\tablenotetext{b}{Membership probability, recalculated for this study.}
\tablenotetext{c}{Binary flag from Paper I: 0, likely single star; 1, likely binary.}
\tablenotetext{d}{Bolometric luminosity from Paper I.}
\tablenotetext{e}{Detection flag: 0, undetected in X ray; 1, detected in X ray.}
\tablenotetext{f}{\LX\ (\LX\ UL) in the 0.5$-$7 keV band for detected (undetected) objects calculated as described in Section~\ref{detections} (\ref{nondetections}).}
\tablecomments{This table is available in its entirety in the electronic edition of the \apj. The first five rows are shown here for guidance regarding its form and content.
}
\end{deluxetable}

\subsection{Calculating \LX\ for Detected Members}\label{detections}
We described our source detection procedure in Paper I. Briefly: we used \texttt{wavdetect} in the \CHANDRA\ Interactive Analysis of Observations \citep[CIAO][]{Fruscione2006} tool and the ACIS Extract  point-source analysis software \citep[AE version 2014feb17;][]{Broos2010}. We found 774 X-ray sources, 278 of which were matched to cluster members. For each source we calculated net counts in the 0.5$-$7 (full), 0.5$-$2.0 (soft), and 2.0$-$7.0 (hard) keV energy bands. 

In Paper I, we converted net count rates into absorbed energy fluxes by calculating the incident flux\footnote{The incident flux is the net counts divided by the mean Auxiliary Response File divided by the exposure time.} in the soft and hard bands, and then multiplying the median photon energy in each band by its incident flux. The absorbed energy flux in the full band is the sum of the energy fluxes of the soft and hard bands. Using the mass bins defined in Section~\ref{masses}, we detected 59 G, 36 K, and 67 M cluster stars. The faintest X-ray emitting M37 member has a mass of 0.26 \Msun.

To compare our results to those for other clusters, we recalculate energy fluxes of cluster members by converting net count rates into unabsorbed 0.5$-$7 keV fluxes using WebPIMMS.\footnote{\url{http://cxc.harvard.edu/toolkit/pimms.jsp}} We assume an atomic neutral hydrogen column density $N_{\mathrm{H}} = 1.26 \times 10^{21}$ cm$^{-2}$, derived from $E$($B-V$) = 0.227 (HA08), $R_V=3.1$, and $N_{\mathrm{H}}$[cm$^{-2}/A_v$] = 1.79 $\times 10^{21}$ \citep{Predehl1995}. We apply a thermal (APEC) model, setting the abundance to 0.4 of solar. This choice of a sub-solar abundance is justified by observations of very active stars, whose coronal abundances range from 0.3$-$0.5 of solar \citep[e.g.,][]{Briggs2003, guedel2004, Telleschi2005, Jeffries2005}.

\begin{figure}
\includegraphics{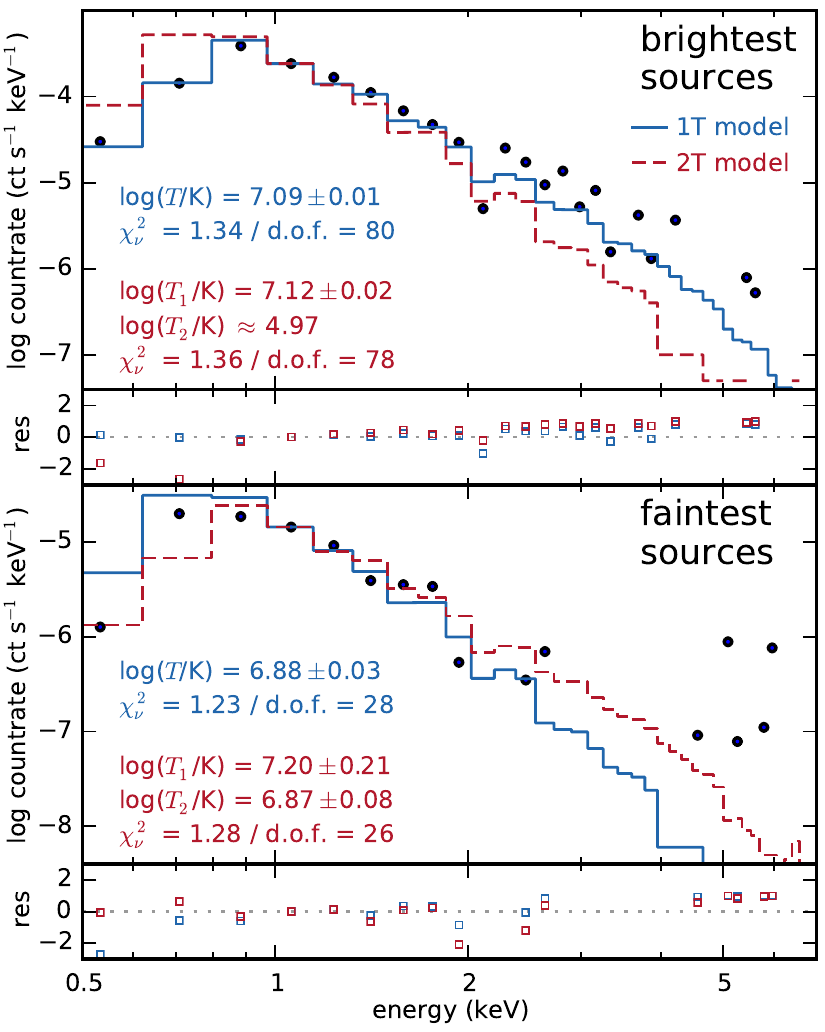}
\caption{Spectral fits for stacks of the 10 brightest  (upper panel) and 40 faintest (lower panel) M37 sources using 1T (blue solid line) and 2T (red dashed line) APEC models, assuming 0.4 solar abundance and $N_{\mathrm{H}}$~=~$1.26 \times 10^{21}$ cm$^{-2}$. The fits are drawn over the binned data (10 counts per bin; black circles). Error bars are too small to show. For each fit, we give the resulting temperature(s) of the APEC model, the reduced chi-squared statistic ($\chi^2_{\nu}$), and the degrees of freedom (d.o.f.). Residuals are shown below each panel, normalized by the stacked spectrum counts. There is no evidence from these fits that a 2T model is required to represent these data.}
\label{fig:specfits}
\end{figure}

We decided to use a one-temperature (1T) model after fitting spectra of the 10 brightest and of the 40 faintest cluster sources\footnote{Stacking is necessary because almost all of our detections have $<<$100 counts.} with 1T and two-temperature (2T) models, and keeping the value of $N_{\mathrm{H}}$ fixed. Figure~\ref{fig:specfits} shows the results: the 2T fits are not a statistical improvement on the 1T fits. In addition, the 2T model does not perform well with the  brightest sources, as the lower of the two temperatures is unreasonably cool (log($T$/K) = 4.97) and likely a result of the fit hitting a hard floor limit. Furthermore, the 2T fit for the faintest sources produces a relatively large uncertainty for the high-temperature component (log($T_2$/K) = 7.20$\pm$0.21). 

A 1T model is therefore an adequate approximation of the plasma temperature for X-ray-emitting members of M37. This is consistent with findings in the literature: typically, 2T fits return plasma temperatures of log($T_1$/K) $\approx$ 6.7 and log($T_2$/K) $\approx$ 7.2 but differ statistically very little from simpler 1T models with a temperature between these two values. This is particularly true for low-count sources \citep[][]{Schmitt1990, Jeffries2005}, and 99\% of our X-ray counterparts to low-mass M37 stars have $<$100 counts. Furthermore, other studies have found that adopting one plasma temperature in the log($T$/K) $\approx$ 6.9$-$7.1 range is adequate for characterizing the underlying differential emission measure in coronae of fairly active stars detected as low-count X-ray sources \citep[][]{Gagne1995, James1997, Jeffries2006, Pillitteri2006}.

We set log($T$/K) = 7 ($kT$ = 0.8617 keV), which is the average of our two 1T models' best-fit temperatures.\footnote{In the Appendix, we discuss further tests we conducted to examine the impact of assuming a constant plasma temperature for our stars regardless of their age.} The median \LX\ of our M37 sample changes by 0.12 dex if we go from adopting log($T$/K)~=~6.9 to log($T$/K)~=~7.1. We are adding $<$1\% in uncertainty to our \LX\ calculations by adopting this single coronal temperature for all sources. 

\subsection{Calculating \LX\ for the Non-Detections}\label{nondetections}
We calculate \LX\ ULs for the 160 undetected low-mass stars with \Pmem\ $\geq$ 0.7 within the field of view of our \CHANDRA\ observation. We follow \citet{Kashyap2010} and define an UL as the maximum \LX\ a source can have without exceeding some detection threshold with a given probability, given a specified background. We use a detection threshold of $10^6$ (equivalent in CIAO to \texttt{wavdetect}'s threshold significance of $10^{-6}$, the value used in the source detection procedure in Paper I) and a false negative probability of 0.5.

To estimate the background of each undetected source, we draw an annulus with \texttt{ds9} centered on its optical counterpart's coordinates. We set the inner radius of the annulus to the size of the point source function (PSF) that encloses 100\% of counts from a point source at that location, and the outer radius to three times that size. We use CIAO's \texttt{dmlist} to find the number of 0.5$-$7~keV counts in the annulus, and \texttt{dmstat} to find the mean exposure time for that region. Combining these two quantities and the area of the annulus, we calculate the background in counts per second per pixel squared, from which we then estimate the background inside the inner radius of the annulus, the actual region of the undetected source.

We convert the resulting UL count rates into unabsorbed X-ray fluxes with WebPIMMS by applying the same model and parameters described in Section~\ref{detections} for members with X-ray detections. In Table~\ref{tbl:M37} we list the \LX\ UL of all undetected stars with \Pmem\ $\geq$ 0.2, although only stars with \Pmem\ $\geq$ 0.7 were included in our analysis.

\subsection{The Impact of ULs on the XLFs}
To define the XLFs and LLFs, we use the Kaplan-Meier (K-M) method as implemented in the \texttt{lifelines} package \citep{Davidson-Pilon2016}, and treat ULs as left-censored data points. We apply Efron's bias correction \citep{Efron1967}, which considers the lowest \LX\ value to be a detection even if it is not. Our M37 sample contains a significant number of X-ray ULs: in the most extreme case, 71\% of the K stars with \Pmem\ $\geq$ 0.7 are undetected. This is concerning because the K-M method is biased when censored data represent a very large fraction of the whole sample \citep[e.g.,][]{Huynh2014}. 

\begin{figure}[h]
\includegraphics{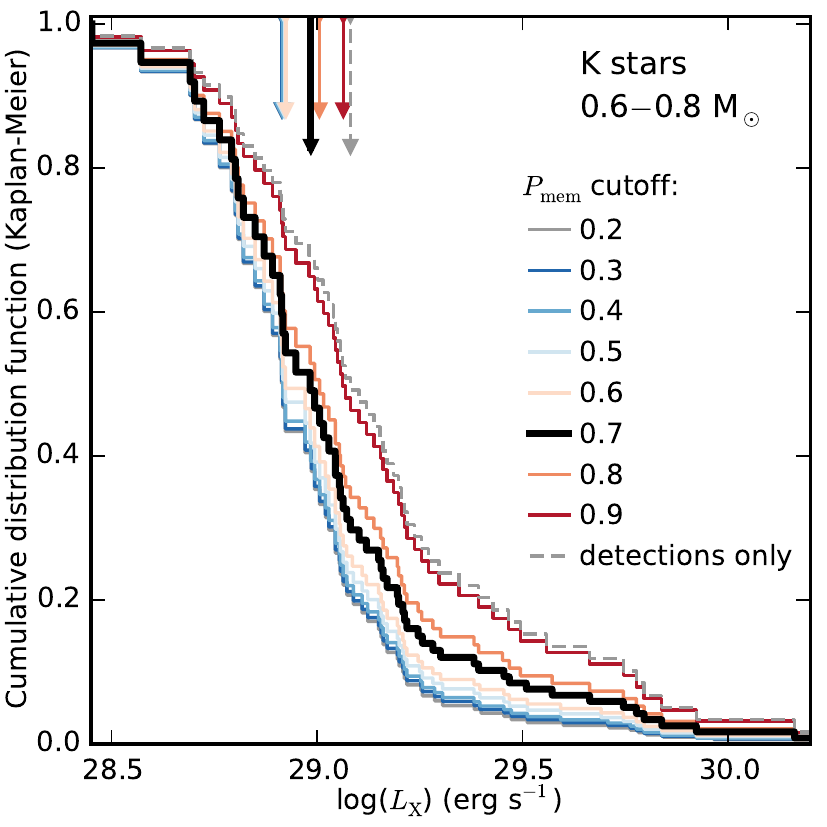}
\caption{K star XLFs for \Pmem\ cutoff values ranging from 0.2 to 0.9 in 0.1 increments. Our adopted cutoff, \Pmem~$\geq 0.7$, is the bold solid black XLF. The K-M solution for the sample including detections only is in dashed gray. The median log(\LX) for each K-M solution is indicated with a vertical arrow.}
\label{fig:XLF}
\end{figure}

To test the impact of ULs, we re-calculate the XLFs after varying the \Pmem\ cutoff, which, since all X-ray detections have \Pmem\ = 1, is equivalent to varying the number of censored data points in the sample. As an example, we show in Figure~\ref{fig:XLF} the impact on the K-star XLF when we vary the \Pmem\ cutoff from 0.2 to 0.9 in 0.1 increments. We also show the XLF we obtain including only detections. Finally, we indicate the median \LX\ from each K-M solution with a vertical arrow.

Having a large number of censored data points biases the XLF toward lower \LX\ values. It appears, however, that implementing a \Pmem\ cutoff $\lapprox$0.8 has little effect on the shape of the XLF. In the extreme case, the difference in median \LX\ between a sample of K stars with \Pmem$~>~0.2$ and one of \Pmem$~>~0.9$ K stars is $\approx$0.2 dex. For G and M stars, the difference in median \LX\ between these two \Pmem\ cutoffs is $\approx$0.1 dex. We conclude that the impact of ULs as a function of \Pmem\ cutoff, and thus of the fraction of ULs in a given mass bin, on the K-M statistics is limited to a shift of at most 0.2 dex.

\section{X-ray Activity at 500 Myr}\label{500}
\begin{figure*}
\begin{center}
\includegraphics{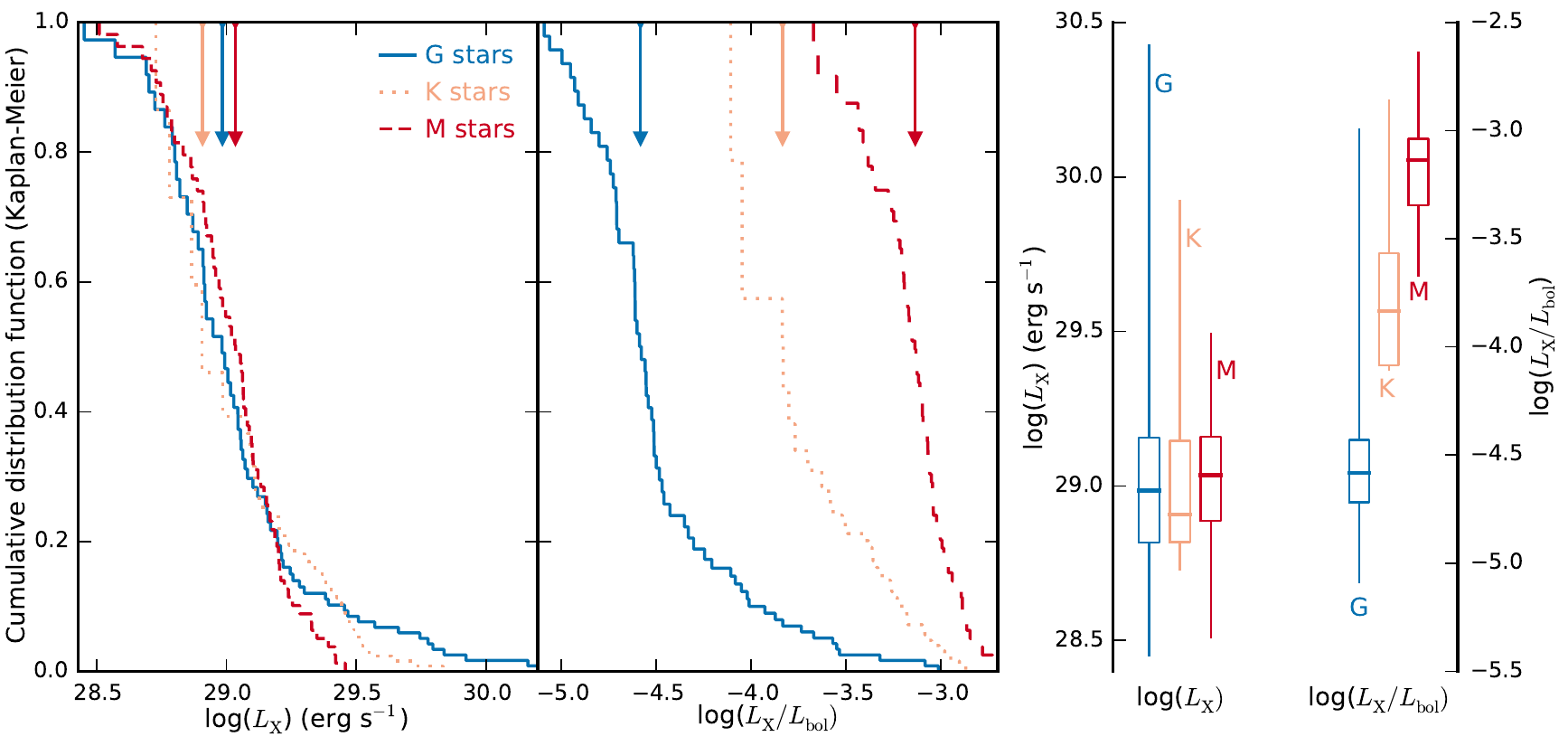}
\caption{K-M estimator for the XLFs (left panel) and LLFs (center panel) of M37 G (solid blue line), K (dotted orange), and M (dashed red) stars. The median value for each K-M solution is indicated with a vertical arrow. To the right of the panels, boxplots for each mass bin extend from the lower to the upper quartile; the whiskers cover the entire data range. The  horizontals lines are the median values.}
\label{fig:M37_LL}
\end{center}
\end{figure*}

The left panel of Figure~\ref{fig:M37_LL} shows the XLFs of stars with \Pmem\ $\geq$ 0.7 for the G (solid blue), K (dotted orange), and M stars (dashed red). The median values are indicated with vertical arrows. G and M stars have the highest median \LX\ values from the K-M solution, log(\LX/\ergs) $\approx$ 29.0, while for K stars it is $\approx$28.9. All three classes have very similar lower$-$upper quartile levels: log(\LX/\ergs) = 28.8$-$29.2 for G and K stars and 28.9$-$29.2 for M stars. This suggests that at an age of 500 Myr, \LX\ is fairly constant for low-mass stars. 

The center panel of Figure ~\ref{fig:M37_LL} compares the LLFs for the three mass bins. Calculating \LL\ allows us to remove the mass dependence of \LX\ and reveal the fraction of the stars' total emission that is in the X ray. The differences in the X-ray contribution to the overall stellar emission are obvious here, with the fraction produced in X rays becoming significantly smaller as mass increases.

Finally, to the right of the panels, boxplots for each mass bin extend from the lower to the upper quartile; the whiskers cover the entire data range, and medians are indicated by a horizontal line inside the boxes. More massive stars clearly have lower intrinsic activity levels than their least massive cousins.

\begin{deluxetable*}{@{}lllcccccccccrrrrr@{}}
\tabletypesize{\scriptsize} 

\tablecaption{Cluster Characteristics And Number of X-ray Detections/Non-Detections \label{tbl:clusters}}

\tablehead{
\colhead{Name} & \colhead{Ref\tablenotemark{a}} & \colhead{Inst.\tablenotemark{b}}  & \colhead{log(\LX)} & \colhead{\ \ } & \multicolumn{2}{c}{G Stars\tablenotemark{d}} & \multicolumn{2}{c}{K Stars\tablenotemark{d}} & \multicolumn{2}{c}{M Stars\tablenotemark{d}} & \colhead{\ \ } & \colhead{log(age)} & \colhead{Distance} & \colhead{log($N_{\mathrm{H}}$)} &
\colhead{$E$($B-V$)} \\
\colhead{} & \colhead{} & \colhead{} & \colhead{Min\tablenotemark{c}} & \colhead{} & \colhead{D} & \colhead{ND} & \colhead{D} & \colhead{ND} & \colhead{D} & \colhead{ND} & \colhead{} & \colhead{(Myr)} & \colhead{(pc)} & \colhead{(cm$^{-2}$)} & \colhead{}\\[-0.1 in]
}

\startdata
ONC & 1 & CA & 27.48 & & 10 & 0 & 7 & 0 & 138 & 0 & & 6.41\tablenotemark{f} & 414.0 & 21.72\tablenotemark{f} & ...\tablenotemark{g} \\
NGC 2547 & 2,3 & RH,X  & 28.83 & & 27 & 0 & 19 & 0 & 44 & 0 & & 7.54 & 407.0 & 20.48 & 0.038 \\
NGC 2516\tablenotemark{e} & 4 & X & 28.53 & & 70 & 56 & 29 & 49 & 96 & 174 & & 8.08 & 385.5 & 20.90 & 0.120\\
Pleiades\tablenotemark{e} & 5,6,7 & RP,RH & 28.00 & & 74 & 23 & 58 & 22 & 88 & 137 & & 8.10 & 136.2 & 20.40 & 0.032\\
NGC 6475 & 8,9,10 & RP,X & 28.64 & & 37 & 0 & 51 & 1 & 6 & 0 & & 8.34 & 302.0 & 20.54 & 0.060\\
M37 & 11 & CA & 28.45 & & 59 & 59 & 36 & 89 & 67 & 12 & & 8.69 & 1490.0 & 21.10 & 0.227\\
Hyades & 12,7 & RP  & 27.50 & & 65 & 12 & 27 & 28 & 54 & 39 & & 8.80 & 47.3\tablenotemark{f} & $<$20.00 & 0.000
\enddata
\tablenotetext{a}{Reference for X-ray data: (1) \citet{Getman2005b}; (2) \citet{Jeffries1998}; (3) \citet{Jeffries2005}; (4) \citet{Pillitteri2006};(5) \citet{stauffer1994a}; (6) \citet{micela1999a}; (7) \citet{stelzer2000}; (8) \citet{Prosser1995}; (9) \citet{James1997}; (10) Obs. ID 0300690101 in 3XMM-DR5 Catalog; (11) Paper I; (12) \citet{stern1995}.}
\tablenotetext{b}{X-ray instrument: CA = \CHANDRA\ ACIS; X = \textit{XMM} EPIC; RP = \ROSAT\ PSPC; RH = \ROSAT\ HRI.}
\tablenotetext{c}{Minimum log(\LX) value detected in \ergs (0.5$-$7.0 keV band).}
\tablenotetext{d}{Number of detections (D) and non-detections (ND). G stars: 0.8$-$1.2 \Msun; K stars: 0.6$-$0.8 \Msun; M stars: 0.1$-$0.6 \Msun.}
\tablenotetext{e}{We combine these two to create the representative $\approx$120-Myr-old cluster used in our analysis.}
\tablenotetext{f}{Mean value for cluster stars.}
\tablenotetext{g}{No reddening value adopted.}

\end{deluxetable*}

\subsection{The Impact of Binaries}
In Paper I we flagged a star as a likely M37 binary if $P_{\mathrm{b}} > P_{\mathrm{s}}$ and $P_{\mathrm{b}} + P_{\mathrm{s}} \geq 0.2$, where $P_{\mathrm{b}}$ was based on height from the main sequence in the ($i$,$g-i$) color-magnitude diagram. Since stars in our M37 sample flagged as likely binary remain unresolved, we are, therefore, potentially overestimating their stellar masses, as we derived masses using the combined photometry of the system. 

Furthermore, some low-mass stars may remain hidden in binaries with a massive companion. This is corroborated by our detection in X rays of 104 high-mass cluster members. X-ray emission from such systems is more likely to come from a low-mass companion \citep[see e.g.,][]{Preibisch2002}, and indeed 39 of the 104 are photometric candidate binaries, suggesting that our sample of M37 low-mass stars is incomplete.

The 44 detected low-mass candidate binaries are also difficult to interpret. The X-ray emission from such systems could potentially come from both components, and so counting each detection as just one source could bias the XLFs and LLFs toward higher luminosities.

We calculate the XLFs by excluding all likely binaries in our sample and compare the results with those in Figure~\ref{fig:M37_LL}. We find very small differences in the K-M solutions of the samples including and excluding likely binaries. In the extreme case, the median log(\LX) of cluster M stars shifts by 0.05 dex when likely binaries are excluded. Furthermore, since potential contamination by binaries is an issue for all XLF studies, and since the binary fraction for open clusters and over time does not vary significantly \citep{Duchene2013}, we simply include binaries in the construction of the XLFs and LLFs of M37 and of all other clusters in our study.

\section{The Evolution of X-ray Activity}\label{longterm}
X-ray activity decreases with time in low-mass main-sequence stars \citep[e.g.,][]{guedel1997, preibisch2005a, giardino2008}. To quantify this evolution, we compare the XLFs and LLFs we obtain for M37 to those for six other open clusters: the ONC \citep[0.1$-$10 Myr,][]{preibisch2005a}, NGC~2547 \citep[35$\pm3$ Myr,][]{Jeffries2005}, NGC~2516 \citep[$120\pm25$ Myr,][]{Silaj2014}, the Pleiades \citep[$125\pm8$ Myr,][]{stauffer1998},  NGC~6475 \citep[220$\pm50$ Myr,][]{Silaj2014}, and the Hyades \citep[625$\pm50$ Myr,][]{Perryman1998}. All six clusters have well-defined membership catalogs that extend to the low-mass end and have been surveyed extensively in the X ray, rendering a meaningful comparison to each other and to M37 possible.

The published \LX\ values from the surveys of these clusters differ in terms of the quoted energy bands and how they were obtained from the instrumental count rates. To homogenize the X-ray data, we use the original count rates of all sources and, as with our M37 sources, apply a 1T APEC model with 0.4 solar abundance and log($T$/K)~=~7 (see further discussion of the assumed $T$ value in the Appendix) to obtain unabsorbed 0.5$-$7 keV fluxes. 

As discussed in Section~\ref{detections}, our choice of plasma temperature is appropriate for low-count X-ray sources of fairly active stars, which describes most stars in these surveys. Furthermore, even though coronal temperatures are found to decrease with stellar age \citep[e.g.,][who found that for solar analogs these temperatures decrease by 0.37 dex between 0.1 and 0.75 Gyr]{Telleschi2005}, the uncertainty introduced by adopting the same temperature for the entire 6$-$625 Myr range is not significant, given the low counts of our sources. Overall, 54\% of the X-ray sources in the surveys considered here have fewer than 100 counts, and 81\% fewer than 500. 

We also found the most up-to-date estimation of distance and reddening for each cluster. We use these to calculate \LX\ and $N_{\mathrm{H}}$, following the reddening-$N_{\mathrm{H}}$ relation of \citet[][]{Bohlin1978}. 

To determine the stars' masses, we first combine $BVRIJHK$ or $ugriz$ photometry, cluster distances, and total absorption in each band\footnote{We use the extinction tables of \citet{schlafly2011} assuming $R_V=3.1$ and adopting $E$($B-V$) values.} to obtain absolute magnitudes. We then apply the absolute magnitude-mass relations of \citet{Kraus2007}; for stars with only $BVRI$ photometry, we use the extended version of the same relations described in Ag\"ueros et al.~(in prep). For the ONC we adopt the stellar masses of \citet{Getman2005b}, which were derived using the pre-main-sequence (PMS) evolutionary tracks of \citet{Siess2000}. 

Finally, we use a linear interpolation between the stellar mass-\Lbol\ values for M37 (see Section \ref{masses}) to estimate \Lbol\ for stars in the other clusters.

\begin{deluxetable*}{lccccccrcc}
\tabletypesize{\scriptsize} 

\tablecaption{Clusters Members Characteristics \label{tbl:clusterstars}}

\tablehead{
\colhead{Name} & \colhead{\ $\alpha$ (J2000)\ } & \colhead{\ $\delta$ (J2000)\ } & \colhead{\ \ Cluster\ \ } & \colhead{\ Mass\tablenotemark{a}\ } & \colhead{log(\Lbol)\tablenotemark{b}} & \colhead{\ Distance\ } & \colhead{$N_{\mathrm{H}}$\tablenotemark{c}} & \colhead{Detection} & \colhead{log(\LX)\tablenotemark{e}} \\  
\colhead{} & \multicolumn{2}{c}{(\arcdeg)} & \colhead{} & \colhead{(\Msun)} & \colhead{\ (\ergs)\ } & \colhead{(pc)} & \colhead{($\times 10^{20}$ cm$^{-2}$)} & \colhead{flag\tablenotemark{d}} & \colhead{(\ergs)}\\[-0.1 in]
}
\startdata
COUP 6  & 83.65928 & -5.40653 & ONC & 0.23 & 33.03 & 414 & 13.18 & 1 & 29.88 \\
COUP 10 & 83.66681 & -5.43448 & ONC & 0.13 & 33.23 & 414 &  1.00 & 1 & 28.68 \\
COUP 14 & 83.67345 & -5.39938 & ONC & 0.13 & 32.92 & 414 &  6.46 & 1 & 28.88 \\
COUP 17 & 83.67930 & -5.33534 & ONC & 0.90 & 33.86 & 414 & 13.18 & 1 & 30.18 \\
COUP 20 & 83.68520 & -5.43502 & ONC & 0.16 & 33.11 & 414 & 48.98 & 1 & 29.44
\enddata
\tablenotetext{a}{Stellar mass derived from $JHK$ or $ugriz$ photometry using the mass-absolute magnitude relation of \citet{Kraus2007} and extended for $BRI$ photometry by Ag\"ueros et al. (in prep).}
\tablenotetext{b}{\Lbol\ derived from a linear interpolation between the stellar mass-\Lbol\ relation in M37 stars, as described in Section~\ref{masses}.}
\tablenotetext{c}{Atomic neutral hydrogen column density; null if assumed to be negligible.}
\tablenotetext{d}{X-ray detection flag: (0) non-detection; (1) detection.}
\tablenotetext{e}{\LX\ (\LX\ UL) in the 0.5$-$7 keV energy band for detected (undetected) objects calculated using the method in Section~\ref{detections}.}
\tablecomments{This table is available in its entirety in the electronic edition of the \apj. The first five rows are shown here for guidance regarding its form and content.}

\end{deluxetable*}

Table~\ref{tbl:clusters} summarizes the basic characteristics of these six clusters and those of M37, and gives for each the number of detections and non-detections in each mass bin. Table~\ref{tbl:clusterstars} lists all low-mass cluster members with their derived stellar masses, \Lbol, distances, $N_{\mathrm{H}}$ values derived from $E$($B-V$), and \LX\ values for detected sources and \LX\ ULs for undetected sources. Below we briefly summarize the X-ray observations for each cluster and the cluster parameters we assumed to perform our analysis.

\subsection{The Comparison Set of Clusters}\label{comparison}
\subsubsection{The ONC}\label{onc}
At 0.1$-$10 Myr in age (we adopt 6 Myr), the ONC serves as an essential young benchmark for studies of the long-term evolution of X-ray activity \citep[e.g.,][]{preibisch2005a, Jeffries2006} because of its well-described membership and extensive X-ray coverage. We therefore include the ONC in our comparison, albeit with two caveats. First, practically all low-mass stars in the ONC are in the PMS phase and therefore still contracting and spinning up. Second, at such young ages low-mass stars are still likely to be surrounded by inner circumstellar disks, which may either obscure or enhance stellar X-ray emission \citep{Bouvier1997, Wolff2004, Preibisch2005b}. Considered all together, ONC stars therefore might not exhibit a clear X-ray rotation-activity relation \citep{Krishnamurthi1997, Feigelson2003}. To select a sample of ONC stars comparable to those in older clusters, we exclude from our analysis stars that show  evidence of having either a circumstellar disk or strong accretion \citep[see sections 8.1 and 8.2 of ][]{Preibisch2005b}.

\citet{Getman2005b} presented a 838 ks \CHANDRA\ observation of the ONC. These authors detected  $>$1600 point sources in the 0.5$-$8~keV band. Following \citet{Preibisch2005b}, we adopt the masses derived by \citet{Getman2005b} using the theoretical PMS evolutionary tracks of \citet{Siess2000}. There are 478 low-mass ONC stars, 155 of which show no evidence for circumstellar disks or strong accretion. We use the published \CHANDRA\ ACIS net counts, exposure times, and $N_{\mathrm{H}}$ values for each of these 155 sources to calculate their \LX. We adopt a distance $d = 414$ pc \citep{Menten2007} for all stars in the cluster.

\citet{Preibisch2005b} reported $>$98\% of low-mass ONC stars as X-ray sources, and there is therefore no need to account for non-detections. The optically faintest cluster member detected in X rays has a mass of 0.10 \Msun.

\subsubsection{NGC~2547}
Although most low-mass stars in the 35-Myr-old cluster NGC~2547 are still in the PMS phase, there is evidence that their inner circumstellar disks have dissipated \citep[e.g.,][]{Jeffries2000, Young2004}. It is thus expected that their X-ray emission be unobstructed and more representative of main-sequence coronae.

NGC~2547 was first observed with \ROSAT\ HRI in the 0.1$-$2.4~keV band for 57.9 ks, resulting in 102 detections of cluster members \citet{Jeffries1998}. The cluster was observed again with \XMM\ in the 0.3$-$3.0~keV band for 49.4 ks; \citet{Jeffries2006} reported 108 detections. In addition, \citet{Jeffries2006} modified the original \ROSAT\ count rates to apply a more sophisticated PSF model for the HRI instrument.

Seventy-two cluster stars are detected in both observations, 36 only by \textit{XMM}, and two only by \ROSAT, for a total of 110 low-mass NGC~2547 stars with X-ray detections. For the \textit{XMM} sources, we use time-weighted mean count rates from their pn, MOS1, and MOS2 count rates. For the \ROSAT\ sources, we use the modified count rates of \citet{Jeffries2006}. For stars detected in both observations, we obtain a combined \LX\ using a weighted average of the two separate \LX\ values. We adopt $d= 407$ pc \citep{Mayne2008}, log($N_{\mathrm{H}}$/cm$^{-2}$) = 20.48 \citep{Jeffries2005}, and $E$($B-V$) = 0.038 \citep{Mayne2008} for all cluster stars.

The vast majority of NGC~2547's low-mass stars are detected. All those with $1.4 < (V-I) < 2.5$ ($\approx$K5$-$M3) are detected, and only a handful of stars near $(V-I)\approx 1.2$ and an increasing number of stars at $(V-I) >$ 2.8 remained undetected. \citet{Jeffries2006} therefore did not account for non-detections. The faintest cluster member detected in X rays has a mass of 0.32 \Msun.

\subsubsection{NGC~2516}
\citet{Pillitteri2006} used two different {\it XMM} pointings totaling 105.7~ks to observe the 120-Myr-old open cluster NGC~2516. These authors detected 258 members (201 low-mass) and calculated 0.3$-$7.9 keV \LX\ ULs for 354 (287 low-mass) that remained undetected.\footnote{NGC~2516 was observed with \ROSAT\ by \citet{Jeffries1997} and \citet{micela2000} and with \CHANDRA\ by \citet{Damiani2003}; given the completeness and much higher sensitivity of the {\it XMM} observation, we opt to use only the latter for simplicity.} The faintest cluster member detected in X rays has a mass of 0.19 \Msun; the same is true for the faintest cluster member with an \LX\ UL measurement. In cases where \citet{Pillitteri2006} matched an X-ray source to more than one cluster star, we assume the X-ray emission to originate from the closest match only.
 
We adopt $d= 385.5$ pc \citep{Terndrup2002}, log($N_{\mathrm{H}}$/cm$^{-2}$) = 20.90 \citep{Pillitteri2006}, and $E$($B-V$) = 0.12 \citep{Dachs1989} for all cluster stars.

\subsubsection{The Pleiades}\label{115Myr}
The Pleiades was surveyed with \ROSAT\ on at least three occasions. \citet{stauffer1994a} first observed the cluster with the PSPC instrument (0.15$-$2.0 keV) using three different pointings for a total of 73.5 ks. These authors detected 176 cluster members, and calculated \LX\ ULs for 62 more members that remained undetected.

\citet{micela1999a} reported several observations of the Pleiades with the \ROSAT\ HRI instrument (0.1$-$2.4~keV) using eight different pointings for a total of 234.7~ks. These authors detected 120 Pleiads, including 15 that were undetected by \citet{stauffer1994a}. \citet{micela1999a} also calculated \LX\ ULs for $\approx$90 members with no previous \LX\ measurements.

Finally, \citet{stelzer2000} calculated 0.15$-$2.0 keV \LX\ for 211 cluster members and \LX\ ULs for 199 undetected ones using 10 publicly available \ROSAT\ PSPC observations with a combined exposure time of 105.9~ks. Sixty-eight of these \LX\ measurements were of cluster members with no previous X-ray detections.

We match these sources to the updated membership catalog of \citet{Covey2016}. We obtain 265 Pleiads with detections and 211 with \LX\ ULs, of which 220 and 182 are low-mass stars, respectively. For stars in more than one X-ray study, we use the weighted average count rate to calculate the \LX. The faintest detected cluster member and the faintest with an upper limit have a mass of 0.12 \Msun. We adopt $d= 136.2$ pc \citep{Melis2014}, log($N_{\mathrm{H}}$/cm$^{-2}$) = 20.4 \citep{micela1999a}, and $E$($B-V$) = 0.032 \citep{An2007} for all cluster stars.

An inspection of the Pleiades and NGC~2516 reveals that the two clusters share several characteristics relevant to our study, including nearly identical XLFs for all mass bins, similar low-mass populations with available X-ray data, and overlapping age estimates. Regarding the latter, we note that both clusters have age estimates spanning the approximate range 65$-$150 Myr. For the Pleiades, several isochronal estimates indicate an age near 125 Myr \citep[e.g.,][]{stauffer1998, david2015, david2016}, but others return ages as young as 75 Myr \citep{steele1993} or as old as 150 Myr \citep{mazzei1989}. Furthermore, lithium-depletion studies indicate ages of 112 Myr \citep{Dahm2015} and 130 Myr \citep{barrado2004}. For NGC~2516, studies of magnetic Ap and Bp stars indicate an age of 120 Myr \citep{Silaj2014}, which is similar to some isochronal age estimates \citep[e.g.,][]{kharchenko2005,lyra2006}, but lower than others \citep[e.g., 140 Myr, 158 Myr,][respectively]{meynet1993, sung2002}.

\begin{figure}
\includegraphics{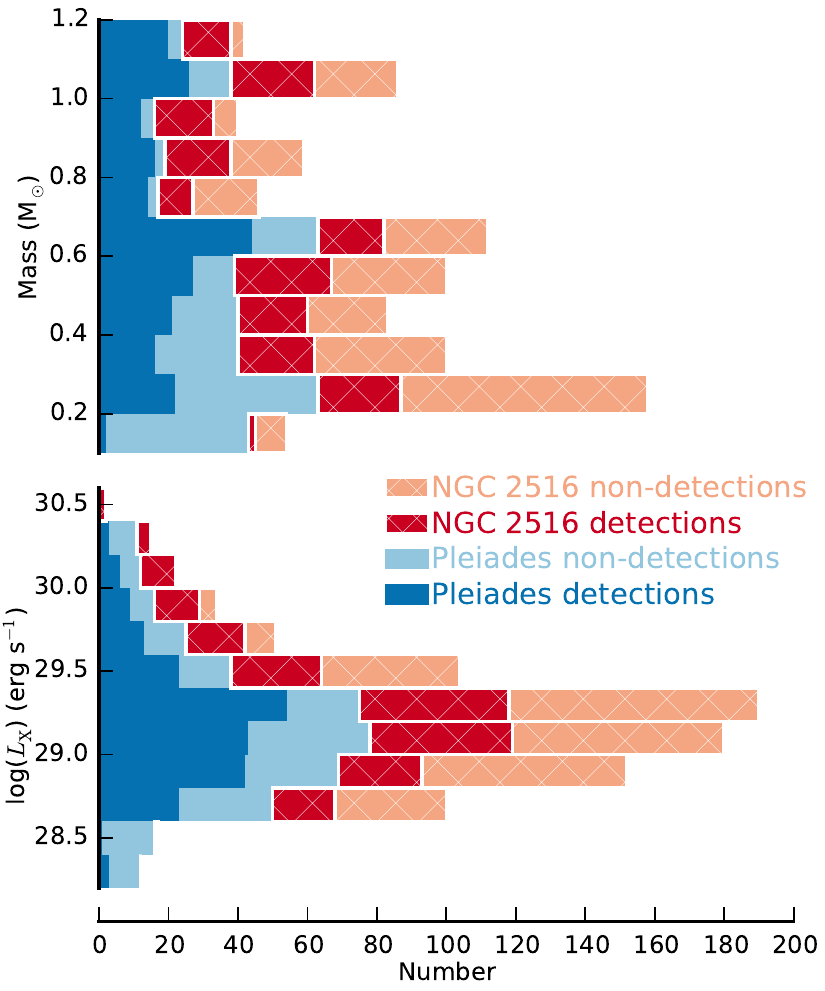}
\caption{Histograms with the number of low-mass detections and non-detections for the Pleiades (light and dark blue) and NGC~2516 (hashed orange and red) as a function of mass (top panel) and log(\LX) (bottom panel). For non-detections, the latter are ULs.}
\label{fig:115hists}
\end{figure}

To simplify our analysis, we combine detections and non-detections from the Pleiades and NGC~2516 to create a single, representative $\approx$120~Myr-old cluster, which we consider to be a reasonable approximate age for stars in the two clusters. Figure~\ref{fig:115hists} shows mass (top) and log(\LX) (bottom) histograms for the resulting cluster.

\subsubsection{NGC~6475}
This 220-Myr-old open cluster was observed by \citet{Prosser1995} with \ROSAT\ PSPC (0.07$-$2.4 keV) using two different pointings for a total of 46.6 ks. These authors found at least one cluster optical counterpart to 129 of their X-ray sources; 24 had two or three counterparts. For the latter group, we assume the X-ray emission to originate from the closest match only.

\citet{James1997} reported a separate 27 ks \ROSAT\ PSPC (0.4$-$2.4 keV) observation of NGC~6475 and the detection of 35 cluster stars, only four of which are not among the \citet{Prosser1995} sources. Neither of the \ROSAT\ surveys reported \LX\ ULs for low-mass stars.

There is also an archival 46 ks {\it XMM} observation (Observation ID 0300690101, PI: R. Pallavicini) in the 3XMM-DR5 catalog of serendipitous X-ray sources \citep{Rosen2015}, for which the Survey Science Center processing pipeline (version 4.1) returned 196 X-ray point sources of good quality (i.e., quality flag = 0). We match 16 of these to low-mass cluster members, all of which are also detected in one or more of the previous X-ray studies. We derive 0.5$-$7.0 keV \LX\ values for these sources using the available 0.2$-$4.5 keV (bands 1$-$4) instrumental count rates.

We search the {\it XMM-Newton} Upper Limit Server\footnote{\url{http://xmm.esac.esa.int/UpperLimitsServer/} for data for the undetected low-mass clusters stars. However, we obtain only one UL for one undetected NGC~6475 star. We derive the 0.5$-$7.0 keV \LX\ UL value for this star from the EPIC-pn 0.2$-$2.0 keV count rate UL estimated by this server.}

In total, there are 133 detected cluster stars, 94 of which are low-mass cluster stars. For stars in more than one X-ray study, we use a weighted average to calculate the \LX. The faintest detected cluster member has a mass of 0.52 \Msun. For all the stars in this cluster we adopt $d= 302$ pc \citep{vanleeuwen2009}, log($N_{\mathrm{H}}$/cm$^{-2}$) = 20.54, and $E$($B-V$) = 0.06 \citep{Robichon1999}. 

\begin{figure*}
\begin{center}
\includegraphics{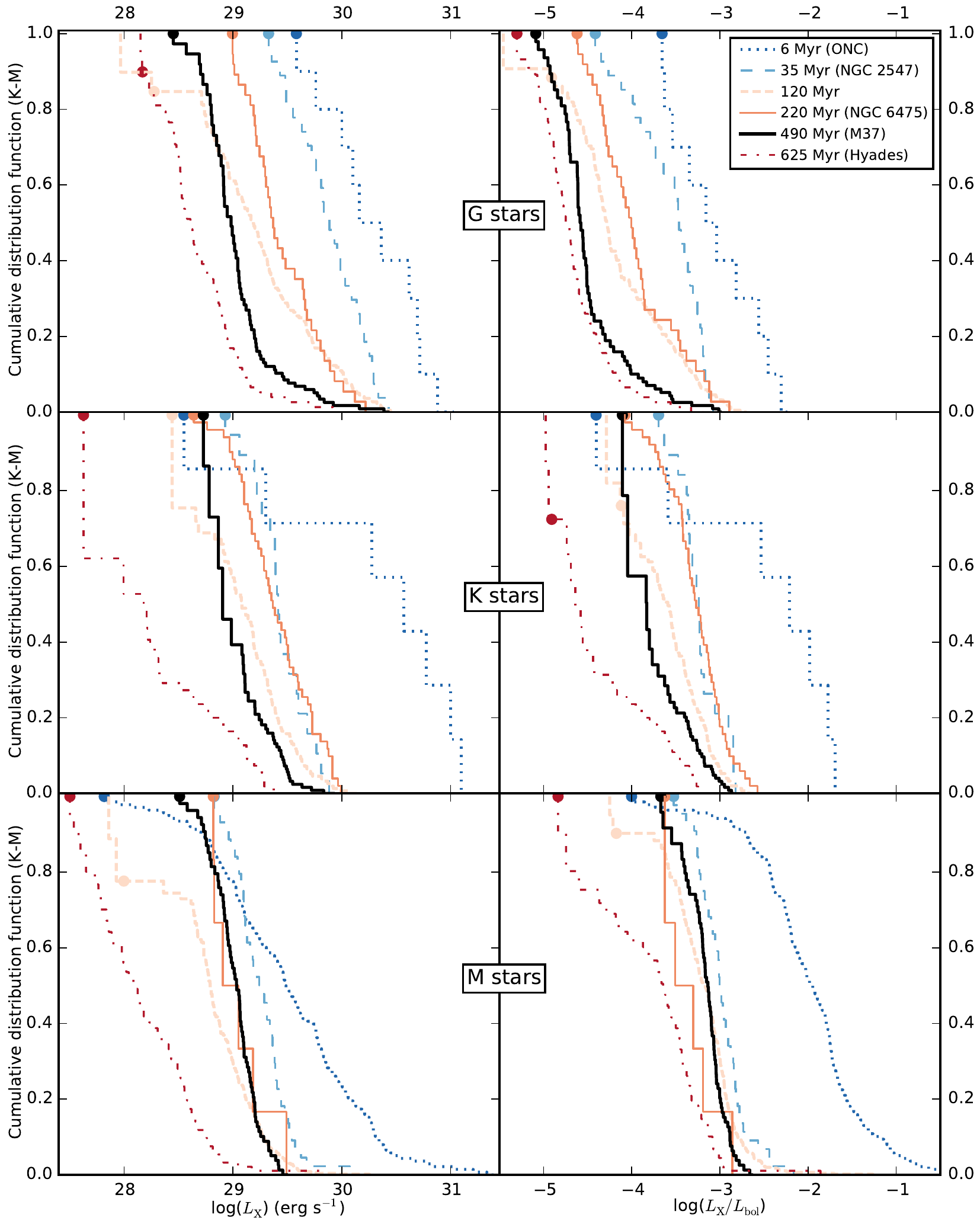}
\caption{K-M estimator for the XLF (left panels) and LLF (right panels) of G, K, and M stars in the ONC ($\approx$6 Myr), NGC~2547 (35 Myr), the merged NGC~2516 and Pleiades clusters (120 Myr), NGC~6475 (220 Myr), M37 (490 Myr), and Hyades (625 Myr). The \LX\ values are all for 0.5$-$7 keV and derived from count rates applying a 1T-plasma model with 0.4 solar abundance and log($T$/K) = 7. The filled circles indicate the minimum \LX\ detected in each cluster at each mass bin.}
\label{fig:XLF_all}
\end{center}
\end{figure*}

\subsubsection{The Hyades}\label{hyades}
\ROSAT\ PSPC (0.1$-$1.8 keV) observations of the 625-Myr-old\footnote{\citet{brandt2015b, brandt2015a} calculated the age of the cluster to be 750$\pm100$ Myr by fitting rotating stellar models to main-sequence turnoff Hyads. However, the quoted uncertainty does not include the additional $\approx$100~Myr systematic uncertainties mentioned by these authors.} Hyades were first obtained by \citet{stern1995}, who examined a $\approx$30$\times 30\deg$ area around the cluster center. These authors detected 188 Hyads and measured \LX\ ULs for 252 that remained undetected. 

\citet{stelzer2001} used publicly available PSPC data to report 0.1$-$2.0 keV \LX\ for 191 Hyads, 36 of which were undetected by \citet{stern1995}. These authors also measured \LX\ ULs for 74 undetected cluster members, 40 of which were not in \citet{stern1995}.

We consolidate the two surveys and match the resulting list of X-ray sources to the Hyades catalog of \citet{Douglas2016}, which combines the catalog of \citet{Goldman2013} with a handful of new Hyads identified from All Sky Automated Survey data (Cargile et al., in prep.). For our analysis, we considered only stars with \Pmem\ $\geq$70\% in the \cite{Douglas2016} catalog.

The result is 178 Hyads with detections and 82 with ULs, of which 143 and 79 are low-mass stars, respectively. The faintest detected Hyad has a mass of 0.13 \Msun, and the faintest Hyad with an \LX\ UL, 0.18 \Msun. For stars in both surveys, we use the weighted average count rate to calculate \LX. We use published star distances and assume negligible reddening and $N_{\mathrm{H}}$. For stars with no available distance, we set $d=46$ pc \citep{vanleeuwen2009}.

\begin{figure*}[!t]
\begin{center}
\includegraphics{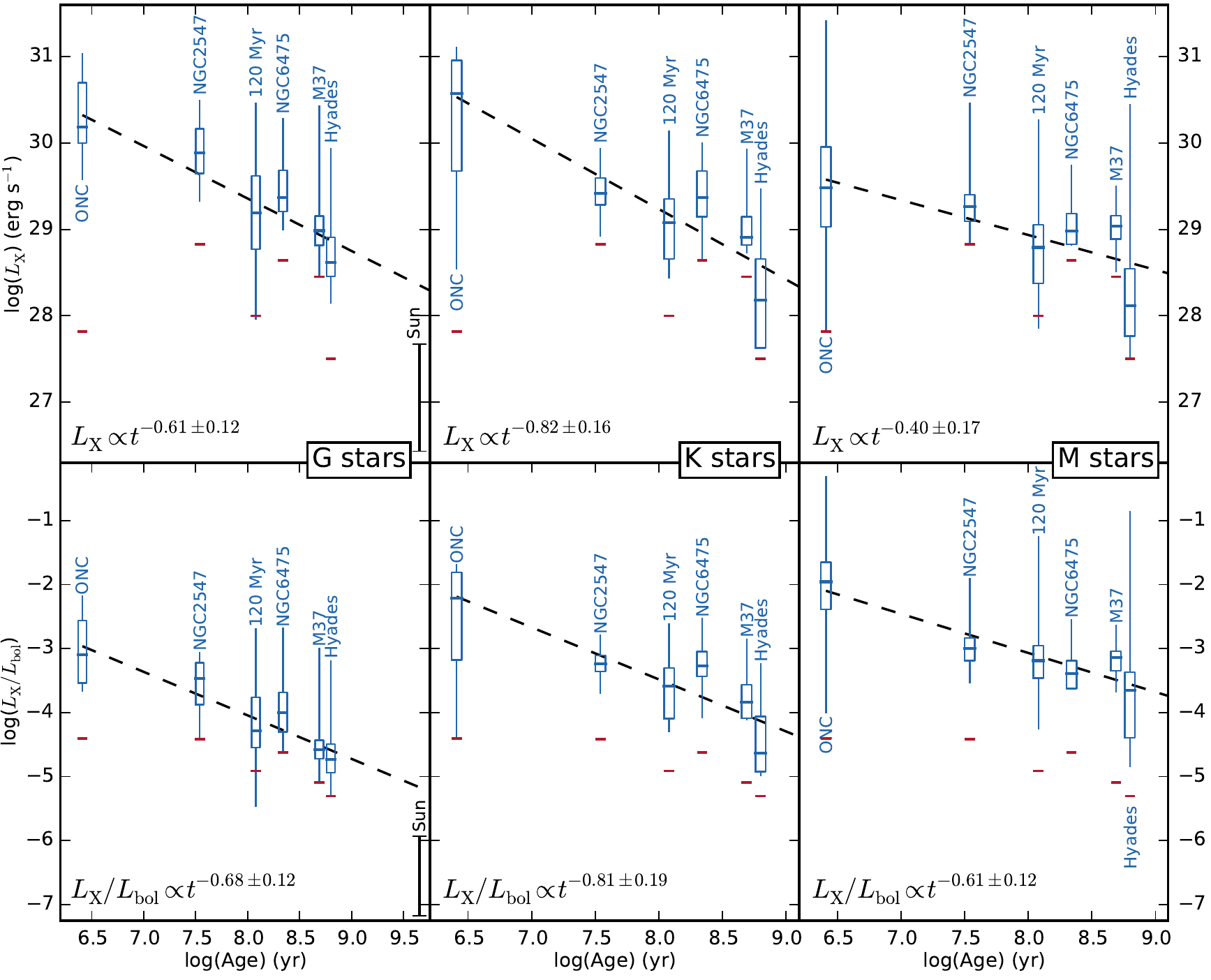}
\caption{The evolution of \LX\ (top panels) and \LL\ (bottom panels) for G, K, and M stars using the K-M estimator results. The representative 120 Myr-old cluster contains stars from NGC~2516 and the Pleiades. Each box extends from the lower to the upper quartile and the whiskers cover the entire data range. The median value is indicated by a horizontal line inside the box. The minimum X-ray value detected in each cluster across the three mass bins is indicated with a red horizontal line. In the G stars panels, the minimum and maximum of the Sun are indicated with black whiskers at 4.5 Gyr. The dashed lines indicate the linear regression analysis in log space (excluding the Sun) for each panel. The resulting dependencies of \LX\ and \LL\ on age are given at the bottom of each panel.}
\label{fig:LXvT}
\end{center}
\end{figure*}

\begin{deluxetable*}{@{}lccccccccccc@{}}
\tabletypesize{\scriptsize} 

\tablecaption{Kaplan-Meier Statistics \label{tbl:results}}

\tablehead{
\colhead{Open} & \multicolumn{5}{c}{log(\LX) (\ergs)\tablenotemark{a}} & \colhead{\ \ \ } & \multicolumn{5}{c}{log(\LL)} \\
\colhead{Cluster} & \colhead{Min} & \colhead{25$^{\rm th}$\tablenotemark{b}} & \colhead{Median} & \colhead{75$^{\rm th}$\tablenotemark{b}} & \colhead{Max} & \colhead{} & \colhead{Min} & \colhead{25$^{\rm th}$\tablenotemark{b}} & \colhead{Median} & \colhead{75$^{\rm th}$\tablenotemark{b}} & \colhead{Max}\\[-0.1 in]
}

\startdata
\multicolumn{12}{c}{\it G stars} \\
ONC & 29.58 & 30.00 & 30.18 & 30.70 & 31.03 & & -3.66 & -3.54 & -3.09 & -2.56 & -2.18\\
NGC 2547\ \ \  & 29.33 & 29.65 & 29.89 & 30.16 & 30.49 & & -4.41 & -3.87 & -3.46 & -3.22 & -3.06\\
120 Myr & 27.97 & 28.77 & 29.19 & 29.62 & 30.46 & & -5.46 & -4.55 & -4.28 & -3.76 & -2.70\\
NGC 6475 & 29.00 & 29.21 & 29.37 & 29.69 & 30.28 & & -4.62 & -4.30 & -4.00 & -3.69 & -2.69\\
M37 & 28.45 & 28.82 & 28.99 & 29.16 & 30.42 & & -5.09 & -4.72 & -4.58 & -4.43 & -3.00\\
Hyades & 28.15 & 28.46 & 28.62 & 28.91 & 29.94 & & -5.30 & -4.94 & -4.73 & -4.49 & -3.19\\
\hline
\\[-0.1 in]
\multicolumn{12}{c}{\it K stars} \\
ONC & 28.55 & 29.67 & 30.57 & 30.96 & 31.10 & & -4.40 & -3.18 & -2.21 & -1.81 & -1.69\\
NGC 2547 & 28.93 & 29.28 & 29.42 & 29.59 & 29.94 & & -3.70 & -3.36 & -3.24 & -3.11 & -2.80\\
120 Myr & 28.44 & 28.66 & 29.08 & 29.35 & 30.14 & & -4.29 & -4.09 & -3.59 & -3.30 & -2.62\\
NGC 6475 & 28.64 & 29.15 & 29.37 & 29.68 & 30.00 & & -4.08 & -3.43 & -3.27 & -3.04 & -2.54\\
M37 & 28.73 & 28.82 & 28.91 & 29.15 & 29.92 & & -4.11 & -4.09 & -3.83 & -3.57 & -2.86\\
Hyades & 27.63 & 27.63 & 28.18 & 28.66 & 29.46 & & -4.98 & -4.92 & -4.63 & -4.06 & -3.24\\
\hline
\\[-0.1 in]
\multicolumn{12}{c}{\it M stars} \\
ONC & 27.82 & 29.03 & 29.48 & 29.95 & 31.41 & & -4.00 & -2.39 & -1.96 & -1.65 & -0.32\\
NGC 2547 & 28.83 & 29.10 & 29.26 & 29.40 & 30.46 & & -3.53 & -3.19 & -3.00 & -2.83 & -1.91\\
120 Myr & 27.86 & 28.37 & 28.79 & 29.05 & 30.26 & & -4.25 & -3.46 & -3.19 & -2.96 & -1.25\\
NGC 6475 & 28.82 & 28.83 & 28.98 & 29.18 & 29.74 & & -3.63 & -3.62 & -3.39 & -3.19 & -2.55\\
M37 & 28.51 & 28.89 & 29.04 & 29.16 & 29.49 & & -3.67 & -3.34 & -3.13 & -3.04 & -2.64\\
Hyades & 27.50 & 27.77 & 28.12 & 28.54 & 30.44 & & -4.84 & -4.40 & -3.65 & -3.37 & -0.86
\enddata
\tablenotetext{a}{0.5$-$7.0 keV band.}
\tablenotetext{b}{Lower and upper quartiles (i.e., 25$^{\rm th}$ and 75$^{\rm th}$ percentiles).}

\end{deluxetable*}

\subsection{Results and Discussion}\label{results}
Figure~\ref{fig:XLF_all} shows the XLF (left panels) and LLF (right panels) K-M solutions for M37 and the six clusters described above, with the Pleiades and NGC~2516 combined into a single 120-Myr-old cluster, for G (upper row), K (middle row), and M stars (bottom row). The XLFs show an overall decrease in X-ray activity spanning approximately two orders of magnitude from the age of the ONC to that of the Hyades. G stars exhibit the most uniform decrease, as each subsequent cluster is approximately half an order of magnitude less luminous than the previous one. The exception is the 120-Myr-old cluster, which is less luminous than this sequence would suggest, both in terms of its XLF and its LLF.

K and M stars evolve less gradually. K stars between 35$-$490 Myr decrease only very slightly in \LX, while the youngest (6 Myr) and oldest (625 Myr) stars appear significantly different from the rest. M stars behave similarly, although the evident differing completeness levels in this mass bin across clusters makes any interpretation of the M-star XLF and LLF difficult (e.g., the least massive star in the ONC is 0.10 \Msun; in NGC~6475, 0.52 \Msun). The varying sensitivity levels of the different cluster surveys hinder a straightforward comparison, as the derived XLFs and LLFs of samples with lower sensitivities imply populations that are more X-ray active than they should be \citep[see e.g.,][]{Feigelson1999, Preibisch2002, Feigelson2003}.

The unexpected behavior of the 120-Myr-old stars was seen by \citet{Jeffries2006}, who found that Pleiads have a similar level of activity as much older Hyads. That study, however, did not include any stars between the ages of the two clusters. Our inclusion of 220- and 490-Myr-old stars makes this odd result stand out. The XLF and LLF curves at 120 Myr are most left-shifted at lower values, suggesting that the explanation may lie in the differences in detection limits between the surveys. 

In Figure~\ref{fig:XLF_all}, the filled circles along each XLF and LLF indicate the faintest X-ray source in each cluster at each mass bin. NGC~6475 and M37 have higher minimum detection values than the 120-Myr-old sample. Furthermore, the significant fraction of ULs in the 120-Myr-old sample is also shifting the XLF toward lower \LX, as we found to be the case for M37 (see Figure~\ref{fig:XLF}).

It is also possible that our XLF at 220 Myr implies a population that is over-luminous relative to reality, as our sample of NGC~6475 stars does not include ULs. If this is true, it would suggest that low-mass stars, and particularly K and M stars, do not decrease in X-ray luminosity by a significant amount between 100 and 500~Myr.

In Figure~\ref{fig:LXvT}, we plot the \LX\ (top panels) and \LL\ (bottom) K-M solutions for each cluster in the three mass bins with boxes and whiskers. Each box extends from the lower to the upper quartile, the horizontal line indicates the median value, and the whiskers span the entire data range for each cluster (Table~\ref{tbl:results} gives these K-M solutions). The faintest X-ray source detected in each cluster across the three mass bins is indicated with a red horizontal line in all panels. This corresponds to the minimum value for each cluster given in Column 4, Table~\ref{tbl:clusters}.

The left panels shows the evolution of G stars, with the X-ray minimum and maximum of the Sun \citep{Peres2000} indicated with black whiskers at 4.5 Gyr. A linear regression analysis in log space using the median values (excluding the Sun) reveals a decrease in \LX\ and \LL\ with time of the form \LX~$\propto~t^b$ and \LL~$\propto~t^c$ (black dashed lines), where $b=-0.61\pm0.12$ and $c=-0.68\pm0.12$. This is much shallower than the slope quoted in Paper I ($b=-1.23\pm0.16$), which included a slightly different set of clusters and no ULs. The correlation coefficients of the fits are $r_b=-0.93$ and $r_c=-0.94$.

\citet{preibisch2005a} found a similar slope $b=-0.76$, but a shallower slope $c\approx-0.5$, for their smaller sample of ONC, Pleiades, Hyades, and nearby field G stars. On the other hand, \citet{Cargile2009} found slopes $b=-0.60\pm0.01$ and $c=-0.64\pm0.41$ for their sample of FG stars from 10 clusters spanning a similar age range. Other studies of solar analogs that extend to ages $>$600 Myr have found a steeper slope $b=-1.5$ \citep{maggio1987, guedel1997, giardino2008}. Our linear regressions extrapolated to the age of the Sun over-predict the X-ray activity of solar-type stars at the solar age by $\approx$0.5 dex in both \LX\ and \LL.

The linear regression analysis for K stars in Figure~\ref{fig:LXvT} (center panels) indicates that \LX\ and \LL\ decrease following the \LX$(t)$ and \LL$(t)$ relations described above with slopes $b=-0.82\pm0.16$ ($r_b=-0.93$) and $c=-0.81\pm0.19$ ($r_c=-0.91$). These are statistically similar to the slopes found for G stars. Although our value for $b$ agrees with that of \citet[][$b=-0.78$]{preibisch2005a} and of \citet[][$b=-0.62\pm0.27$]{Cargile2009} for their samples of K stars, our $c$ slope is steeper than these authors' ($c\approx-0.5$ and $c=-0.34\pm0.32$, respectively).

The relations found for M stars (right panels) have slopes $b=-0.40\pm0.17$ ($r_b=-0.76$) and $c=-0.61\pm0.12$ ($r_c=-0.93$), the former being statistically different from that of K stars. Otherwise, these slopes are all within one standard deviation of the results for the other mass bins. \citet{preibisch2005a} found a steeper slope $b=-0.69$, while \citet{Cargile2009} had a shallower slope of $b=-0.30\pm0.21$, for their M stars. Analogously to the case of G stars beyond the age-range we studied, \citet{Stelzer2013} found a steeper slope $b=-1.10\pm0.02$ for a sample of M0$-$M3 stars in the solar neighborhood covering the ages 0.1$-$3 Gyr. 

In \LL\ space, however, our slope $c$ is significantly steeper than the slopes found by \citet{preibisch2005a} ($c\approx-0.3$) and \citet{Cargile2009} ($c=-0.08\pm0.26$). In the framework of the rotation-activity relation, the similar decay in X-ray activity we find for M and GK stars may suggest that even if a different braking mechanism operates for fully convective stars (spectral types $\approx$M6 and later), the decay in coronal heating nonetheless occurs at the same rate as in stars with radiative cores. The evident incompleteness at the lowest masses in several of the clusters studied here prevents us, however, from making any strong claims about the X-ray evolution of these stars.

Unlike in Figure~\ref{fig:XLF_all}, where the 120-Myr-old stars appear under-luminous in X rays for their age compared to NGC~6475 and M37, Figure~\ref{fig:LXvT} and Table~\ref{tbl:results} show that these stars do not deviate much from the general trend observed from 6 to 625 Myr. It is therefore very likely that the somewhat unexpected shape of the XLF at 120 Myr is mostly a construct of the lower sensitivity and completeness of the NGC~6475 and M37 X-ray surveys relative to those of the Pleiades and NGC~2516.

Is a linear fit adequate for the full age range we consider? Clearly, we do not have enough points in age space to explore fitting broken power laws with different slopes for different age ranges. Instead, we redo the linear regression analysis by excluding open clusters that either should or appear to behave differently from the others. For instance, even though we exclude ONC stars with circumstellar disks or strong accretion (see Section~\ref{onc}), X-ray emission from this cluster may still not be truly comparable to that of older clusters, as its low-mass PMC members may be spinning up on their way to the main sequence. At the other end of our age range, Hyades stars may have already passed an evolutionary threshold beyond which X-ray activity decreases at a faster pace, as seen in studies of older clusters such as the $>$1-Gyr-old cluster NGC~752 \citep{giardino2008}.

Linear fits in log space for \LX\ and \LL\ excluding the ONC result in slightly steeper slopes for all mass bins except for K stars in \LX\ space and M stars in \LL\ space; the latter results in a slope shallower by a factor of two. However, the new slopes lie well within one standard deviation of the original fits, and they all have worse (i.e., closer to zero) $r_b$ and $r_c$ values than the original fits. This suggests that the overall X-ray activity of such young stars does not significantly deviate from the general decaying trend observed in the first 600 Myr of their lives. 

On the other hand, linear fits excluding the Hyades result in slightly shallower slopes for all three mass bins. Again, the new slopes lie within one standard deviation of the original fits, and they all have very similar $r_b$ and $r_c$ values. Therefore, we see no evidence that the behavior of stellar coronae has changed significantly by the time low-mass stars reach the age of the Hyades. However, the scarcity of well-characterized $>$1-Gyr-old open clusters prevents us from conducting a more detailed study of a hypothetical distinct late evolutionary stage of stellar activity beyond 625~Myr.

Finally, we examine whether assuming the same $10^7$~K coronal temperature for all stars in the age range 6$-$625 Myr significantly impact our results. For the youngest stars in our sample (i.e., ONC members), doubling the representative plasma temperature decreases the derived \LX\ values by 0.03$-$0.77 dex; the larger the value of $N_{\mathrm{H}}$, the larger the decrease. For the oldest stars in our sample (i.e., Hyads), cutting the temperature in half decreases the derived \LX\ values by 0.17 dex (see Appendix for further discussion). The effect of these decreased \LX\ values at $\approx$6 Myr and 625~Myr on the age-activity relations of Figure~\ref{fig:LXvT} is inconsequential: the largest change in fitted parameters is the slope $c$ for M stars, which becomes slightly steeper ($-0.69\pm0.18$). We are therefore confident that our decision to chose one representative plasma temperature is reasonable for this analysis. 

\section{Conclusion}\label{summary}
We use a 440.5 ks \CHANDRA\ observation of M37 to characterize the XLFs and LLFs of $<$1.2 \Msun\ stars at 490 Myr. In Paper I we detected 162 such stars; here we add ULs for 160 cluster stars that were undetected in our original observations. At 490 Myr, these G, K, and M stars exhibit similar levels of X-ray activity (median log(\LX) $\approx$ 29.0 \ergs) with similar statistical spreads (28.8$-$29.2 as the lower$-$upper quartiles). In \LL\ space, on the other hand, we find that more massive stars produce a smaller fraction of their overall energy output in the X~ray, with median \LL\ values of $-$4.73, $-$3.83, and $-$3.13 for G, K, and M stars, respectively. 

To characterize the evolution of X-ray activity, we compare the M37 XLFs and LLFs to those of other open clusters in the approximate age range 6$-$625 Myr: the ONC, NGC~2547, NGC~2516, the Pleiades,  NGC~6475, and the Hyades. We homogenize the X-ray data from different surveys by converting published count rates into energy fluxes using the same model and parameter values: a 1T APEC model with 0.4 solar abundance and log($T$/K) = 7. As in other studies, we find that this choice of temperature is an adequate approximation for active stars detected as low-count X-ray sources.

We use up-to-date cluster reddening and distance measurements to calculate $N_{\mathrm{H}}$ and \LX, respectively. We obtain the masses for members of all clusters by applying the same photometric color-mass relation, except for stars in the ONC, for which we use previously published masses. Finally, we calculate \Lbol\ from a linear interpolation between mass-\Lbol\ values we derive for M37 stars.

The XLFs and LLFs of stars in the approximate age range 6$-$625 Myr shift toward lower luminosities over time. G and K stars decrease in X-ray activity by almost two orders of magnitude, as indicated by their median \LX\ and \LL\ values, whereas M stars decrease by about 1.5 orders of magnitude. 

The XLFs and LLFs of stars at 120 Myr (NGC~2516 and the Pleiades combined) appear under-luminous compared to that of younger and older clusters. This is likely to be the result of the large number of ULs included in these XLFs and LLFs and of the relatively higher detection sensitivity of the X-ray surveys of the Pleiades.

Conversely, the XLFs and LLFs of NGC~6475 may be the anomalous ones, as the lack of ULs in this cluster may make it seem over-luminous for its age. These results highlight the difficulty in comparing X-ray surveys with different sensitivities and completeness. Most clusters included here deserve deeper X-ray studies to fully characterize the evolution of activity in low-mass stars.

The decay rate over the approximate age range 6$-$625 Myr is well described by a single linear fit in log space. The decay follows the relation \LX~$\propto~t^b$, with $b=-0.61\pm0.12$ for G stars, $-0.82\pm0.16$ for K stars, and $-0.40\pm17$ for M stars. In \LL\ space, the decay follows \LL~$\propto~t^c$, with $c=-0.68\pm0.12$ for G stars, $-0.81\pm0.19$ for K stars, and $-0.61\pm0.12$ for M stars. 

The incompleteness of the M star data in several clusters prevents us from making any strong conclusions, but our results are incompatible with the paradigm of slower rotational decay rates for M stars compared to G or K stars. Based on our data, the difference in braking mechanisms between fully convective mid- to late-type M stars and GK stars with radiative cores manifests itself only marginally in the coronal heating process. 

The decay rates in \LX\ and \LL\ for G stars over-predict the X-ray activity of the Sun. This, together with existing results from solar analogs with $t>1$ Gyr showing $b=-1.5$, suggests that at older ages than those sampled here the age-activity relation may be significantly steeper for solar-mass stars. Excluding the Hyades from our linear fits results in slightly shallower slopes but similar statistical fits for all three mass bins. Thus, if there is a more rapid decline in X-ray activity at older ages, it appears to take place beyond 625 Myr.

\acknowledgments We thank Vinay Kashyap for his insights into UL computations and K-M estimations, and Jeremy Drake for his comments on constructing XLFs. We thank the anonymous referee for comments that helped improve the paper. We thank the SAO Pre-doctoral Program for hosting A.N.~for three months at the Harvard Smithsonian Center for Astrophysics in Cambridge, MA. This research has made use of data obtained from the 3XMM XMM-Newton serendipitous source catalogue compiled by the 10 institutes of the XMM-Newton Survey Science Centre selected by ESA. Support for this work was provided by NASA through \CHANDRA\ Award Number G02-13025A issued by the Chandra X-ray Observatory Center, which is operated by the Smithsonian Astrophysical Observatory for and on behalf of NASA under contract NAS8-03060. M.A.A.\ acknowledges support provided by the NSF through grant AST-1255419.

\setlength{\baselineskip}{0.6\baselineskip}
\bibliography{references}
\setlength{\baselineskip}{1.667\baselineskip}

\appendix\label{Tparam}
\section{Is it appropriate to use $T = 10^7$ K to model coronae for low-mass stars ranging from 6 to 600 Myr in age?}
In Section~\ref{detections}, we model the X-ray emission for all the stars in our sample using a single coronal temperature of $T=10^7$ K. This implies minimal evolution of this temperature for low-mass stars from 6 to 625 Myr, which is somewhat surprising: higher $T$ values may be more appropriate for the youngest stars and lower $T$ values for the oldest. Here, we investigate the effect of varying the coronal temperature: we cut our benchmark $T$ in half for the oldest stars in our sample (i.e., Hyads), double it for the youngest ones (i.e., ONC stars), and keep $T=10^7$ K for all other stars.

We find that cutting $T$ in half for the Hyads leads to the derived \LX\ values decreasing by 0.17 dex. The \LX\ values also decrease when we double $T$ for the ONC stars,  albeit as a function of $N_{\mathrm{H}}$, which ranges from 20 < log($N_{\mathrm{H}}$/cm$^{-2}$) < 23 for this cluster (for the Hyades, $N_{\mathrm{H}}$ is negligible). For log($N_{\mathrm{H}}$/cm$^{-2}$) values of 20, 21, 22, and 23, \LX\ decreases by 0.03, 0.06, 0.30, and 0.77 dex, respectively. In our ONC sample, 42\% of stars have log($N_{\mathrm{H}}$/cm$^{-2}$) < 21, 91\% have log($N_{\mathrm{H}}$/cm$^{-2}$) < 22, and 100\% have log($N_{\mathrm{H}}$/cm$^{-2}$) < 23, so that for the bulk of these stars the decrease is $<$0.30 dex.

We then test how much the age-activity linear regression fits described in Section~\ref{results} change if we use the halved and doubled $T$ values for Hyads and ONC stars. Table~\ref{tbl:diffT} compares the original fits based on an assumed $T=10^7$ K for all stars in our sample to fits using the new \LX\ values derived for Hyads and ONC stars. We find that assuming different coronal temperatures for our two bookend clusters does not have a significant impact on our results. To first approximation, $T=10^7$ K is an appropriate choice for the plasma temperature for the stars in our sample over the age range 6$-$625 Myr.

\begin{deluxetable}{@{}cccc@{}}
\tabletypesize{\scriptsize}
\tablecaption{Impact of Using Different Coronal $T$ Values on Linear Regression Fits to the Age-Activity Relation \label{tbl:diffT}}

\tablehead{
\colhead{Coronal} & \colhead{G Stars} & \colhead{K Stars} & \colhead{M Stars} \\
\colhead{Temperature} & \colhead{Slope} & \colhead{Slope} & \colhead{Slope} \\[-0.1 in]
}
\startdata
\multicolumn{4}{c}{\it \LX\ v.~$t$} \\
$T=10^7$ K & $-0.61\pm0.12$ & $-0.82\pm0.16$ & $-0.40\pm0.17$\\
Alternative $T$\tablenotemark{a} & $-0.61\pm0.15$ & $-0.83\pm0.23$ & $-0.46\pm0.26$\\[0.1 in]
\multicolumn{4}{c}{\it \LL\ v.~$t$} \\
$T=10^7$ K & $-0.68\pm0.12$ & $-0.81\pm0.19$ & $-0.61\pm0.12$\\
Alternative $T$\tablenotemark{a} & $-0.61\pm0.15$ & $-0.80\pm0.23$ & $-0.69\pm0.18$
\enddata
\tablenotetext{a}{$T=5 \times 10^6$ K for Hyads, $T=2\times 10^7$ K for ONC stars, and $T=10^7$ K for all other stars in our sample.}
\end{deluxetable}

\end{document}